\documentclass[pra,prl,twocolumn,superscriptaddress,floatfix,showpacs,10pt]{revtex4}
\usepackage{graphicx}
\usepackage{amssymb}
\usepackage{verbatim}

\begin{document}

\title{Measuring the parity of an $N$-qubit state}

\author{B. Zeng}
\affiliation{Department of Physics, Massachusetts Institute of
Technology, Cambridge, MA 02139, USA}

\author{D.L. Zhou}
\affiliation{School of Physics, Georgia Institute of Technology,
Atlanta, Georgia 30332, USA} \affiliation{Institute of Theoretical
Physics, The Chinese Academy of Sciences, Beijing 100080, China}

\author{L. You}
\affiliation{School of Physics, Georgia Institute of Technology,
Atlanta, Georgia 30332, USA} \affiliation{Institute of Theoretical
Physics, The Chinese Academy of Sciences, Beijing 100080, China}

\date{\today}

\begin{abstract}
We present a scheme for a projective measurement of
the parity operator $P_z=\prod_{i=1}^N \sigma_z^{(i)}$
of $N$-qubits. Our protocol uses a single ancillary qubit,
or a probe qubit, and involves manipulations
of the total spin of the $N$ qubits without requiring
individual addressing. We illustrate our protocol
in terms of an experimental implementation with atomic ions
in a two-zone linear Paul trap, and further discuss its
extensions to several more general cases.
\end{abstract}


\maketitle

Quantum measurement, the only approach to gain information on
the state of a quantum system, is an essential ingredient of
the quantum theory \cite{Cav,Bra,Hel,Gio}. In quantum
computation and information, systems of many qubits are
prepared, processed, and measured. Quantum measurements on
a system of many qubits can be simply classified into single
qubit measurements and quantum collective measurements, or
measurements on many qubits as a whole at the same time.
A typical example of quantum collective measurement is the Bell
state measurement on two qubits as in quantum teleportation,
where the four entangled and orthogonal
Bell states have to be distinguished \cite{Ben,Bou,Kim}.
Although quantum collective measurements are more difficult to
implement in laboratory systems than single qubit measurements,
they are a necessary element for useful applications of
quantum computations, especially for the implementation of
quantum error corrections. The theoretical foundation of
quantum collective measurements is well established \cite{Hel,Nie}.
In principle, they can be
implemented by a series of unitary transformations and single
qubit measurements with the aid of auxiliary (ancillary) qubits \cite{Nie}.
For specific laboratory systems, however, how to implement a given
quantum collective measurement \textit{efficiently} remains a
challenging task.

In this article, we suggest a scheme for a collective measurement
of the parity operator $P_z$ of $N$ qubits. The efficiency of
our protocol is due to: 1) only one
ancillary qubit (probe qubit) is introduced;
and 2) its overall complexity does not explicitly
depend on the number of qubits $N$.

The parity operator $P_z\equiv\prod_{i=1}^{N}\sigma_z^{(i)}$
of $N$ qubits is important for quantum information science.
It is frequently invoked in the foundations of
quantum mechanics for its direct connections to nonlocal quantum correlations.
Its measurement is also important for
highly precise sensing schemes \cite{Win1,Sac,Win2}, e.g. in the realization
of Heisenberg limit atomic metrology with a maximally entangled state
of $N$ qubits. The simplest scheme for measuring $P_z$
involves projective measurements on
individual qubits ($|0\rangle$ or $|1\rangle$),
destroying coherence among multi-qubit states
within the same parity subspace along the way. An instructive method was proposed recently
by Fenner \textit{et al.} \cite{Fen}, which uses pairwise Heisenberg
interaction and $N+1$ auxiliary qubits.

We start with an intuitive method for measuring $P_z$ as described
in the following steps: (i) We introduce a probe qubit (denoted by
$0$) prepared initially in state $|0^{z}\rangle_0$, the eigenstate
of $\sigma_z^{(0)}$ with eigenvalue $+1$; (ii) We perform
successively $N$ control-NOT (C-NOT) gates between the i-th qubit
(control) and the probe qubit, where $i$ runs from $1$ to $N$;
(iii) We measure the probe qubit in the basis states of
$\sigma_z^{(0)}$. If the measurement result is $+1$, the parity of
the $N$ qubit state is $+1$, and the $N$ qubit system is projected
into the subspace with $+1$ parity; Otherwise the parity is $-1$,
and the state of $N$ qubits is projected into the subspace with
$-1$ parity. The complexity of such a scheme comes mainly from
step (ii), where the probe qubit is required to interact
sequentially with qubit-$1$ to qubit-$N$, to effect the $N$ C-NOT
gates.

\begin{figure}[htbp]
\includegraphics[width=2.4 in]{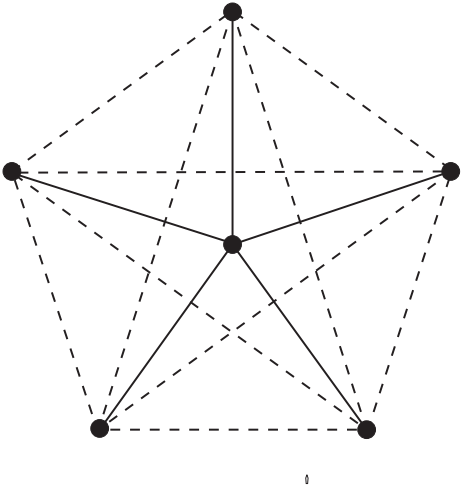}
\caption{A graph illustrating our protocol for measuring
the parity of a $N=5$ qubit state. The ancillary probe qubit
is located at the center, and the parity of the five surrounding
system qubits is to be measured.} \label{fig1}
\end{figure}

The scheme we propose is based on the following idea. We intend to
perform the $N$ C-NOT gates not sequentially, but in only one step.
In fact, we execute $N$ two-qubit phase gates
[see Eq. (\ref{defphasgat})] instead because it is more convenient
to scale them up due to the invariance under interchange of the two
qubits. Our protocol is conveniently illustrated in terms of
the graph as in Fig. \ref{fig1}. In the first step,
we perform phase gates to every qubit pairs from
the $N+1$ qubits, i.e. the probe qubit plus the $N$ system qubits.
In the second step we perform phase gates to every qubit pairs
of the $N$ system qubits. In Fig. \ref{fig1},
a black solid dot denotes a qubit (for $N=5$).
A phase gate between two qubits is represented by a line that
connects the two dots. In the first step, a total of $C_{N+1}^{2}$
lines are needed; while $C_{N}^2$ lines result from the
second step. Because the inverse of a phase gate is
itself, the net effect of the above two steps is a graph with $N
(=C_{N+1}^{2}-C_{N}^{2})$ solid lines connecting the ancillary probe qubit
and the $N$ system qubits as in Fig. \ref{fig1}.
The dashed lines denote the actions of repeated phase gates, or null events.

We now describe the four steps of our protocol:

(i) The probe qubit is initialized to
\begin{equation}
|0^{x}\rangle_0=\frac {|0^{z}\rangle_0+|1^{z}\rangle_0}
{\sqrt{2}}.
\end{equation}

(ii) Phase gates are applied to all pairs of qubits, the probe qubit and
the $N$ system qubits, through the unitary transformation
\begin{equation}
U^{(0,1,\cdots,N)}(N+1,J_z)=\prod_{(ij)}U^{(ij)},\;\;
i,j\in\{0,1,2,\cdots,N\}
\label{uoN}
\end{equation}
where $(ij)$ denotes the $i$-th and $j$-th qubit pair,
on which the action of the phase gate $U^{(ij)}$ is defined as
\begin{equation}
U^{(ij)} |a^{z}\rangle_i |b^{z}\rangle_j = (-1)^{ab}
|a^{z}\rangle_i |b^{z}\rangle_j,\;\; a,b\in\{0,1\}.
\label{defphasgat}
\end{equation}
Several earlier theoretical studies have shown that the unitary transformation
$U^{(0,1,\cdots,N)}$ can be realized (in one step) by the following Hamiltonian
\cite{Mol,You},
\begin{equation}
H(N+1,J_z)=\chi\left( \frac {J_z^2} {2}-\frac {N} {2} J_z+\frac
{N^2-1} {8}\right), \label{jz2}
\end{equation}
with $J_z\equiv\sum_{i=0}^{N} {\sigma_z^{(i)}}/{2}$ the
$z$ component of the total spin of the $N+1$ qubits.
In fact, this is easy to see as
$$U^{(0,1,\cdots,N)}(N+1,J_z)=e^{-iH(N+1,J_z) {\pi}/{\chi}}.$$

(iii) The unitary transformation
$U^{(1,2,\cdots,N)}(N,J_z)$ is applied to the N system qubits.

(iv) The measurement of $\sigma_x^{(0)}$ on the probe qubit
is carried out.
If the result is $+1$ (collapsing the probe to state $|0^x\rangle_0$),
the parity of the $N$ qubit state is $+1$, and the
state of the $N$ system qubits is projected into the subspace of parity $+1$;
Otherwise the parity is $-1$, and the state of
$N$ qubits is projected into the subspace of parity $-1$.

Because different phase gates $U^{(ij)}$s commute with each other,
the step (iii) of the above protocol can be moved to either before
the step (ii) or after the step (iv). The unitary transformation
$U^{(1,2,\cdots,N)}(N,J_z)$ is also known to be capable of
generating specific cluster states, e.g. an $N$-qubit GHZ state,
when the $N$-qubit state is initialized to a product state
$\prod_{i=1}^N |0^{x}\rangle_i$ \cite{Bri,Got}. Certain
error-correcting codes can be realized with cluster states, or
more generally with graph states \cite{Schl}. As mentioned before,
the Hamiltonian (\ref{jz2}) can be realized in various two state
systems, e.g., quantum degenerate atoms \cite{You}, trapped atomic
ions \cite{Mol}, cavity QED systems \cite{Pel}, and solid state
Josephson junctions \cite{Shn,wp}.

We now prove that our protocol indeed corresponds to a projective
measurement of $P_z$ for an $N$-qubit system in a pure state
\begin{equation}
|\psi\rangle_{12\cdots
N}=\sum_{a_1,a_2,\cdots,a_N=0}^{1}\left(c_{a_1a_2\cdots
a_N}\prod_{i=1}^N|{a_i}^{z}\rangle_i\right).
\label{ps}
\end{equation}
From the definition of the two qubit phase gate (\ref{defphasgat}),
it is easy to find that
\begin{eqnarray}
U^{(0;1,2,\cdots,N)}&\equiv&
U^{(1,2,\cdots,N)}(N,J_z)U^{(0,1,\cdots,N)}(N+1,J_z)\nonumber\\
&=&\prod_{i=1}^N U^{(0i)}.
\end{eqnarray}
Then the state of the whole system before the step (iv) is simply
\begin{eqnarray}
&&U^{(0;1,2,\cdots,N)}|0^{x}\rangle_0|\psi\rangle_{12\cdots
N}\nonumber\\
&&={|0^{x}\rangle_0}|\psi^{e}\rangle_{12\cdots
N}+{|1^{x}\rangle_0}|\psi^{o}\rangle_{12\cdots N},\label{entp}
\end{eqnarray}
where
\begin{equation}
|1^{x}\rangle_0=\frac {|0^{z}\rangle_0-|1^{z}\rangle_0}
{\sqrt{2}},
\end{equation}
and
\begin{eqnarray}
|\psi^{e}\rangle_{12\cdots N}
&=&\frac {1+P_z} {2} |\psi\rangle_{12\cdots N}=\delta_{P_z,+1}
|\psi\rangle_{12\cdots N},
\label{evenstat}\\
|\psi^{o}\rangle_{12\cdots N}&=&\frac {1-P_z} {2}
|\psi\rangle_{12\cdots N}=\delta_{P_z,-1} |\psi\rangle_{12\cdots
N},\label{oddstat}
\end{eqnarray}
are respectively the even and odd parity parts of the state (\ref{ps}).
In deriving Eqs. (\ref{entp}), (\ref{evenstat}), and
(\ref{oddstat}), we have made use of the eigen-equation for the
parity operator
\begin{equation}
P_z \prod_{i=1}^N|{a_i}^{z}\rangle_i=(-1)^{\sum_{i=1}^N
a_i}\prod_{i=1}^N|{a_i}^{z}\rangle_i.
\end{equation}
Through the steps (ii) and (iii) we entangle the eigenstates of
$\sigma_x^{(0)}$ for the probe qubit with the parity eigenstates
of the $N$-qubit system, as illustrated by solid lines in Fig.
\ref{fig1}. A measurement of $\sigma_x^{(0)}$ on the probe qubit
thus leads to a projective measurement of the parity operator
$P_z$ on the $N$ system qubits.

Similarly, the probe qubit can be initialized
to other states, e.g. the $|1^{x}\rangle_0$. We then find analogously
\begin{eqnarray}
&&U^{(0;1,2,\cdots,N)}|1^{x}\rangle_0|\psi\rangle_{12\cdots
N}\nonumber\\
&&={|0^{x}\rangle_0}|\psi^{o}\rangle_{12\cdots
N}+{|1^{x}\rangle_0}|\psi^{e}\rangle_{12\cdots N},\label{entm}
\end{eqnarray}
i.e. entanglement is again established, and the parity of the $N$
system qubits is revealed from measuring the probe qubit.
Clearly, our protocol and the above proof also apply to
mixed states of $N$ qubits.

Compared to the intuitive scheme first discussed,
our protocol possesses two main advantages:
(1) The Hamiltonian $H(N,J_z)$ is
a function of the total qubit number $N$ and the $z$ component of
the total spin, thus addressing of individual qubits is not required,
or explicitly, the complexity of
our protocol is independent of $N$;
(2) The ancillary probe qubit has the same
interaction properties as the $N$ system qubits,
so one type of qubits can be used for experimental implementations.
We can simply consider one of the $N+1$ qubits as the probe.

We now contrast our protocol with other schemes as well as adopted
experimental approaches for collective measurements of parity.
The eigenvalues for the parity $P_z$ of $N$ qubits take
two alternative values, $+1$ or $-1$. Their corresponding
subspaces are thus $2^{N-1}$ fold degenerate, and can be
further characterized by the number of $1$s, or
$N^{(-)}=\sum_{i=1}^N ({1-\sigma_z^{(i)}})/{2}$.
Our projective measurement protocol collapses the N-qubit state
into the degenerate subspace of a fixed parity, but maintains
quantum coherence between states of different $N^{(-)}$ because
$\left[N^{(-)},P_z\right]=0$. Previous schemes \cite{Win1,Win2}
and approaches adopted by several recent experiments \cite{Sac,Mit,Zha},
on the other hand, measure $P_z$ by counting the number of qubits
in state $|1^z\rangle$, or evaluating $N^{(-)}$ from results
of individual qubit measurements.
We note that $P_z$ is $+1$ (or
$-1$) when $N^{(-)}$ is even (or odd). In these schemes
of measuring $P_z$ through $N^{(-)}$, coherence between states
with different $N^{(-)}$ but with the same parity $P_z$
is completely destroyed, thus they cannot be used for
quantum error correction.

\begin{figure}[htbp]
\includegraphics[width=3. in]{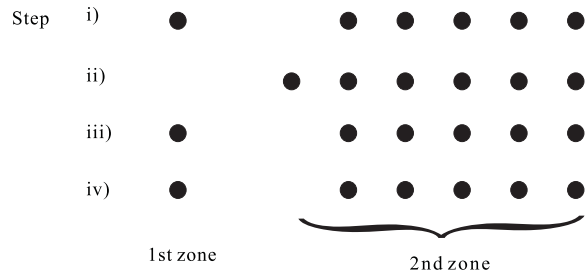}
\caption{From top to bottom, the configurations of ions
in each of the four steps for our protocol implemented in
a two-zone radio
frequency linear Paul trap.} \label{fig2}
\end{figure}

For a simple implementation of our protocol, we consider
atomic ions in a two-zone radio frequency linear Paul trap \cite{Win3}.
The ion configurations during each of the above four steps
are illustrated in Fig. \ref{fig2}.
In steps (i), (iii),
 and (iv), the ancillary probe ion is in the first zone, while the $N$
system ions are in the second zone; In step (ii), all the $N+1$ ions are in
 the second zone, where the unitary
 transformations $e^{i\chi J_x^2}$ and $e^{i(\alpha J_x+\beta J_y)}$
 can be effected experimentally \cite{Win3}. Thus one can use the equalities
 \begin{eqnarray}
 e^{i\chi J_z^2}&=&e^{i\frac {\pi} {2} J_y}e^{i\chi J_x^2}e^{-i\frac {\pi} {2} J_y},\\
e^{i\alpha J_z}&=&e^{i\frac {\pi} {2} J_y}e^{i\alpha
J_x}e^{-i\frac {\pi} {2} J_y},
\end{eqnarray}
to realize the required unitary transformations (\ref{uoN})
in steps (ii) and (iii). It seems such capabilities
already exist in current experimental systems \cite{Win3}.
As already pointed out, steps in the second zone only
involve the manipulation of the total spin, no individual ion qubit
addressing is required.

In the following, we extend our protocol to more general quantum
collective measurements.
For instance, let's consider the measurement of a
general element of the Pauli group $G_N$
\begin{equation}
{\cal P}=P^{(1)}_x \otimes P^{(2)}_y \otimes P^{(3)}_z \otimes
P^{(4)}_e,\label{genpar}
\end{equation}
where $P^{(\alpha)}_{\mu}=\prod_{i=1}^{N_{\alpha}}
\sigma_{\mu}^{(\alpha,i)}$, ($\alpha=1,2,3,4$; and $\mu=x,y,z,e$),
 with $\sigma_e$ the identity operator in a two-dimensional
Hilbert space, and the total qubit number $N=N_1+N_2+N_3+N_4$.
This extension is not only necessary for the completeness of
our theory, it is also required for implementing quantum error
corrections. Our strategy for quantum collective measurement
of ${\cal P}$ now proceeds as follows.
First, the $N$ qubits are grouped into four
sets of $N_1$, $N_2$, $N_3$, and $N_4$ qubits as in (\ref{genpar}).
With only one ancillary probe qubit ($0$), ${\cal P}$ is measured
in the end after successive applications of steps i), ii), and iii)
of our protocols to $P^{(1)}_x$, $P^{(2)}_y$, and $P^{(3)}_z$.
The probe qubit measurement thus makes this scheme a projective measurement
protocol for ${\cal P}$.

To accomplish the above strategy for measuring ${\cal P}$, we
extend our protocol for the parity operator $P_z$ to collective
operators $P_x$ and $P_y$, by making use of state rotations, or
equivalently cyclic coordinate transformations $x\rightarrow y
\rightarrow z \rightarrow x$. Such transformations enable
measuring $P_x$ and $P_y$ in similar fashions to the protocol for
measuring $P_z$ as discussed above. Explicitly, one first prepares
the probe qubit in the state $|0^y\rangle_0$, and then effects the
unitary transformations needed for the measurement of $P_x$; Next
a unitary transformation for the probe qubit is carried out:
$\{|0^{y}\rangle_0\rightarrow |0^{z}\rangle_0,\
|1^{y}\rangle_0\rightarrow |1^{z}\rangle_0\}$, and $P_y$ can be
measured accordingly with appropriate unitary transformations; One
then performs a final unitary transformation on the probe qubit:
$\{|0^{z}\rangle_0\rightarrow |0^{x}\rangle_0,\
i|1^{z}\rangle_0\rightarrow |1^{x}\rangle_0\}$, and measures the
parity operator $P_z$ using our protocol; In the end, a projective
measurement of $\sigma_x^{(0)}$ for the probe qubit reveals the
value of ${\cal P}$.

Before conclusion, we briefly generalize our protocol to the case of
$d$-level systems (qudits). A natural generalization of
two state Pauli matrices
$\sigma_z$ and $\sigma_x$ are operators $Z$ and $X$,
satisfying $XZ=qZX$ with $q\equiv \exp(i2\pi/d)$
\cite{Sun,Zzb,Wey,Sch}. The parity of $N$ qudits can
then be defined as
$P_z\equiv \prod_{i=1}^{N} Z^{(i)}$, whose eigenvalues take
on values
$q^{i},\; (i=0, 1, \cdots, d-1)$. If we define the appropriate
two-qudit phase gate as
\begin{equation}
U^{(ij)} |a^{z}\rangle_i |b^{z}\rangle_j = q^{ab} |a^{z}\rangle_i
|b^{z}\rangle_j,\;\; a,b\in\{0,1,\cdots, d-1\},
\end{equation}
our above protocol for qubits can be directly generalized to
qudits. For example, Equation (\ref{entp}) now becomes
\begin{eqnarray}
U^{(0;1,2,\cdots,N)}|0^{x}\rangle_0|\psi\rangle_{12\cdots
N}=\sum_{n=0}^{d-1}\delta_{P_z,q^n}{|n^{x}\rangle_0}|\psi\rangle_{12\cdots
N}.\hskip 12pt \label{entpd}
\end{eqnarray}
The measurement of $X^{(0)}$ therefore gives rise to a
projective measurement of $P_z$ for the $N$ qudits.

In summary, we have proposed a theoretical protocol for a
projective measurement of the parity $P_z$ of an $N$-qubit state,
making use of a moving probe qubit and collective spin
interactions. Provided symmetric spin interactions can be
engineered without addressing individual spins, the complexity of
our protocol becomes independent of the qubit number $N$. We have
suggested an experimental implementation of our protocol with ions
in a two-zone linear Paul trap. It seems that all necessary steps
are within the limits of current laboratory capabilities
\cite{Win3}. Compared to previous schemes for measuring the same
parity operator, our protocol makes use of nonlinear interactions
to effect a quantum collective measurement in terms of one
ancillary qubit, rather than individual projective measurements on
all qubits. Thus, our projective scheme maintains the quantum
coherence in the subspace of the parity of $N$ qubits, and can be
applied to quantum error correction. Finally, we have generalized
our protocol to quantum collective measurements of any elements of
the Pauli group for $N$ qubits, as well as to systems of qudits.

This work is supported by NSF, NASA, and NSFC.

\end{document}